\def\gs{\mathrel{\raise1.16pt\hbox{$>$}\kern-7.0pt
\lower3.06pt\hbox{{$\scriptstyle \sim$}}}}
\def\ls{\mathrel{\raise1.16pt\hbox{$<$}\kern-7.0pt
\lower3.06pt\hbox{{$\scriptstyle \sim$}}}}

\documentstyle[editedvolume,numreferences,epsfig]{crckapb10}

\begin{opening}
\title{Searching (the) FIRST radio arcs near ACO clusters}
\subtitle{}

\author{H. ANDERNACH}
\institute{Dpto.\,de Astronom\'\i a, IFUG, Apdo.\,Postal 144, Guanajuato, Mexico}
\author{A.G. GUBANOV}
\institute{SAO RAS, Astron.\,Inst., St.-Petersburg State Univ., Russia
198904}
\author{O.B. SLEE}
\institute{Australia Telescope National Facility, Epping, NSW 2121, Australia}

\end{opening}

\runningtitle{Search for Radio Arcs in ACO clusters}

\begin{document} 

\vspace*{-7.5cm}
\begin{footnotesize}\baselineskip 10 pt\noindent
Proc. {\it Observational Cosmology with the new Radio Surveys}, Tenerife,
Spain, Jan.\,13--15, 1997 \\
eds.~~M.\,Bremer, N.\,Jackson \& I.\,P\'erez-Fournon, Kluwer Acad.\,Press,
in press
\end{footnotesize}
\vspace*{6.8cm}

\begin{abstract}
Gravitational lensing (GL) of distant radio sources by galaxy clusters 
should produce radio arc(let)s.
We extracted radio sources from the FIRST survey near Abell cluster cores 
and found their radio position angles to be uniformly distributed with
respect to the cluster centres.  This result holds even when we restrict 
the sample to the richest or most centrally condensed clusters, and to
sources with high S/N and large axial ratio. 
Our failure to detect GL with statistical methods may be due to poor cluster 
centre positions. We did not find convincing candidates for arcs either.
Our result agrees with theoretical estimates predicting
that surveys much deeper than FIRST are required to detect the effect.
This is in apparent conflict with the detection of such an effect claimed 
by Bagchi \& Kapahi~(1995). 
\end{abstract}

\section{Motivation and Theoretical Expectations}

The first gravitational lens (GL) was found upon identification of
the ``double'' quasar 0957+561 \cite{Wal79}.
Later, the first lensed images of extended optical objects were discovered 
as the ``giant arcs'' \cite{LP86}, and since then CCD imaging of lensed 
galaxies behind galaxy clusters has turned into an ``industry''. This
allowed both mapping of the clusters' gravitational potential  
as well as unprecedented insight into high-z galaxies, due to the
clusters acting as natural gravitational ``telescopes'' \cite{SEF92}.
Up to now, no {\it radio}~ arcs in clusters have yet been identified,
although radio searches have yielded more than half of all 
confirmed cases of ``strong lensing'' \cite{KeeKoc96}.
Compared to optical galaxies, radio sources are {\it rare}~
(their sky densities are comparable for S$_{1.5}\sim$0.1\,mJy and 
B$_{J}\sim$24.0 mag),
and they are more extended and complex, making it difficult to 
distinguish lensed from unlensed sources.

We have the highest chance to see radio arcs in massive, centrally 
condensed clusters at z$\sim$0.15--0.6. These tend to be very rich, 
X-ray luminous, and of morphologically ``early'' type BM\,I or RS cD. 
In the absence of large-area samples of distant (z$\gs$0.3)
clusters, we chose our lens candidates from the ACO \cite{ACO89} catalog
and the sources from FIRST \cite{FIRST}, the highest-resolution (5.4$''$)
large-area radio survey available, with a high source density 
($\approx$3\,10$^5$ sr$^{-1}$ for S$\gs$1\,mJy at 1.5\,GHz).

The simplest estimate of N$_{l}$, the number of radio sources expected to be
strongly lensed by clusters, may be derived from the formula for the 
Einstein radius $\theta_{E}$,
the characteristic angular separation from the lens-observer axis within
which giant arcs and multiple imaging occur:\\[1ex]
\hspace*{35mm} $\theta{_E}^2$ = 4 G M$_{E}$ c$^{-2}$ d$_{ls}$ d$_l^{-1}$ d$_s^{-1}$~,\\[1ex]
where G is the gravitational constant, M$_{E}$ is the
cluster mass within a radius d$_{l}$~$\theta{_E}$ from the cluster core
(M$_{E}\sim$10$^{14}$M$_{\odot}$),    
c is the speed of light, d$_{l}$ and d$_{s}$  are the angular-diameter 
distances to the lens and to the source, and d$_{ls}$ is the distance between
lens and source. A rough estimate of N$_{l}$ is then  \\[1ex]
\hspace*{12mm} N$_{l}~~\approx$~~(G\,M$_E$\,H$_0$\,c$^{-3}$) $\langle$z$_i^{-1}\rangle$ N$_{cl}$ 
A$_{s}~~\approx$~~1.6 10$^{-9}\langle$z$_i^{-1}\rangle$ N$_{cl}$ A$_s$ , \\[1ex]
where 
H$_0$=100\,h$_{100}$\,km/s/Mpc is Hubble's constant,
N$_{cl}$ the number of ``suitable'' clusters and
z$_i$ their redshifts, $\langle..\rangle$ is the average,
and A$_s$ the surface density of ``suitable'' radio sources.
Taking the 60 richest (R$\ge$2) ACO clusters with z$\ge$0.1 covered by
FIRST, we have $\langle$z$_{i}^{-1}\rangle\approx$6. The 585 sources 
within 1 Abell radius (1\,R$_{\rm A}$= 1.5 h$^{-1}_{100}$ Mpc)
imply a surface density of $\sim$5 10$^5$ sr$^{-1}$.
Thus we have N$_{l}\sim$0.3\,k, where k$\approx$0.3 is the estimated 
fraction of {\it distant}\, FIRST sources with compact details, i.e.
``suitable'' for lensing.
We obtain N$_{l}\sim$0.1, or at most N$_{l}\sim$1 if we include 
all 373 ACO clusters covered by FIRST.
For a {\it statistical} detection of arc(let)s in ACO clusters we need a survey 
reaching a limiting flux of $\ls$0.1\,mJy.


Other, more detailed estimates \cite{WuHa93} agree that any giant 
radio arcs in ACO clusters should be below the flux limit of FIRST.
Despite of this, a statistical GL effect was claimed to have
been detected in a VLA snapshot survey of 46 distant (D$\ge$5 or
z$\gs$0.1), cD-type ACO clusters at 1.5\,GHz and HPBW=30$''$ \cite{BK95}. 
Within d$_{\rm A}$=0.25 of 28 clusters (where d$_{\rm A}$ is the projected 
distance from the cluster centre in units of R$_{\rm A}$) 40 resolved sources of 
deconvolved size 8$''$--35$''$ were found.
Let $\phi$ be the acute angle between the radio source major axis and 
the line joining the radio centroid with the optical cluster centre,
i.e. $\phi$=0$^{\circ}$ for ``radial'' and 90$^{\circ}$ for 
``tangential'' orientation.
The distribution of this ``orientation angle'' $\phi$ of the 40 sources
peaked near 65$^{\circ}$ ($\phi_{\rm med}$=61$^{\circ}\pm5^{\circ}$). 
This preference for tangential source orientations was noticable out 
to d$_{\rm A}$=0.7, i.e.\,far beyond typical Einstein radii, 
and was interpreted by \cite{BK95} as likely due to lensing.
If confirmed, this would suggest that the above estimates were pessimistic
in that massive, extended dark haloes in clusters are much more frequent 
than hitherto assumed.

We tried to measure this effect more accurately and cross-correlated the 
ACO catalog with the 96\,May\,28 version of FIRST 
(138,665 sources; $\alpha$,$\delta_J$=
[6.6$^{\rm h}$..17.6$^{\rm h}$; +28.2$^{\circ}$...+42.0$^{\circ}$].
We also looked for individual candidate radio arcs in FIRST.
FIRST's angular resolution is five times better, but with 
$\sigma\sim$0.15 mJy its sensitivity for sources $\gs$10$''$
is worse than that of \cite{BK95}. 
Among the 373 ACO clusters covered by FIRST there are 28 cD-clusters. Nine of 
these clusters contain 29 extended FIRST sources with d$_{\rm A}<$0.25,
with deconvolved sizes from 2$''$ to 14$''$.

\section{Methods and Results}

We find $\sim$9200 FIRST sources (of any shape and flux) within 1\,R$_{\rm A}$
of all ACO clusters. Their centre positions were taken from \cite{ACO89}, 
and Abell radii from \cite{ATS95}.  
Within 0.55\,R$_{\rm A}$ of 334 different clusters we find $\sim$700 sources 
more than randomly expected, i.e.~one or two genuine cluster 
members per Abell cluster, depending on the amount of complex sources 
broken up into components in the FIRST catalog.

The FIRST catalog lists deconvolved major ($\theta_a$) and 
minor ($\theta_b$) axes and a radio position angle (PA)
for {\it every}\, source. Both $\theta_a$ and $\theta_b$ may be
negative whenever the effect of noise yielded a source fit smaller than
the beam.  We discarded 33\% of all FIRST sources:
26\% had $\theta_{a}<$1$''$ or else $\theta_{a}$=$\theta_{b}$, 
i.e. were either too small or ``too round'' to trust their orientation.
Another 7\% had $\theta_{b}<$0 and $|\theta_{b}|>|\theta_{a}|$, i.e. 
PA was ill-defined.
For each FIRST source located within 1\,R$_{\rm A}$ and having a 
significant radio PA we calculated
its orientation angle $\phi$ as defined in Section~1.

Before fine-tuning our sample for the search for radio arcs we 
selected those FIRST sources with peak flux S$_{p}\ge$2\,mJy and axial ratio 
$\epsilon\equiv\theta_{a}/\theta_{b}\ge$1.5 in 
clusters with distance class D$\ge$5. We then compared 
the distribution of $\phi$ for the 1065 sources with d$_{\rm A}\le$0.75 
in 269 different clusters, and those of the 258 sources with d$_{\rm A}\le$0.25
in 130 clusters. Both samples were
split into three equal parts, the first one at
d$_{\rm A}=$0.33 and 0.56 (Fig.\,1a), and the second one 
at d$_{\rm A}=$0.10 and 0.18 (Fig.\,1b).
Histograms for the inner (solid lines), the middle (dashed lines) and
the outer annuli (dotted lines) around cluster centres
show no trend for $\phi$ to peak at higher values for sources closer to 
the cluster cores (i.e.\,smaller d$_{\rm A}$),
contrary to findings of \cite{BK95}.
A Kruskal-Wallis test shows that the distributions have probabilities 
of 25\% (Fig.\,1a) and 20\% (Fig.\,1b) 
of being drawn from the same population of orientation angles.
A Kolmogorov test shows that uniformity cannot be rejected at
significance levels of $\gs$10\,\% ($\sim$5\,\%  for 0.1$<d_{\rm A}<$0.18).

If the clusters of our sample are similar to those with known optical arcs, 
then their Einstein radii must be of order 30$''$ to 60$''$. 
Inaccurate Abell centres in ACO of $\sim$2$'$ rms \cite{ACO89} could 
potentially invalidate the above analysis and will create 
random errors in $\phi$, large enough to smooth out a real peak in a 
$\phi$ distribution of sources with d$_{\rm A}\le$0.25.
In order to minimise this source of error, we defined a subsample 
of 65 sources with d$_{\rm A}\le$0.25
in clusters of type BM$<$II. The latter have central dominant galaxies 
and, we assume, both better-defined ACO centres and a higher probability 
of showing GL. For this sample, and a sub-sample of 25 sources 
(S$_{p}\ge$2\,mJy, $\epsilon\ge$1.5) we obtain the highest values 
of $\langle\phi\rangle$ and $\phi_{\rm med}$ of $\sim$50$^{\circ}$ (Fig.\,1c).
A Kolmogorov test is unable to reject uniformity at 
significance levels of 26 resp. 16\%.

Errors in ACO's centre coordinates should be relatively less
important for nearer clusters (at least in units of R$_{\rm A}$).
However, for a sample of 177 FIRST sources 
(S$_{p}\ge$2\,mJy, $\epsilon\ge$1.5) with d$_{\rm A}<$0.25 in 79 ACO 
clusters of distance class 4 and 5,
we found $\langle\phi\rangle$=43$^{\circ}$ and $\phi_{\rm med}$=39$^{\circ}$.
A Kolmogorov test gives a significance level of 5\,\%
for rejection of uniformity.
Thus we find no statistical evidence for radio arcs in Abell clusters in
the FIRST survey, neither by selecting clusters by their distance,
richness, or morphological type, nor by selecting radio sources by their
proximity to the cluster centre, their flux or their ellipticity.

\begin{figure}
\hspace*{-9mm}
\vspace*{-4mm}
\mbox{
\epsfig{file=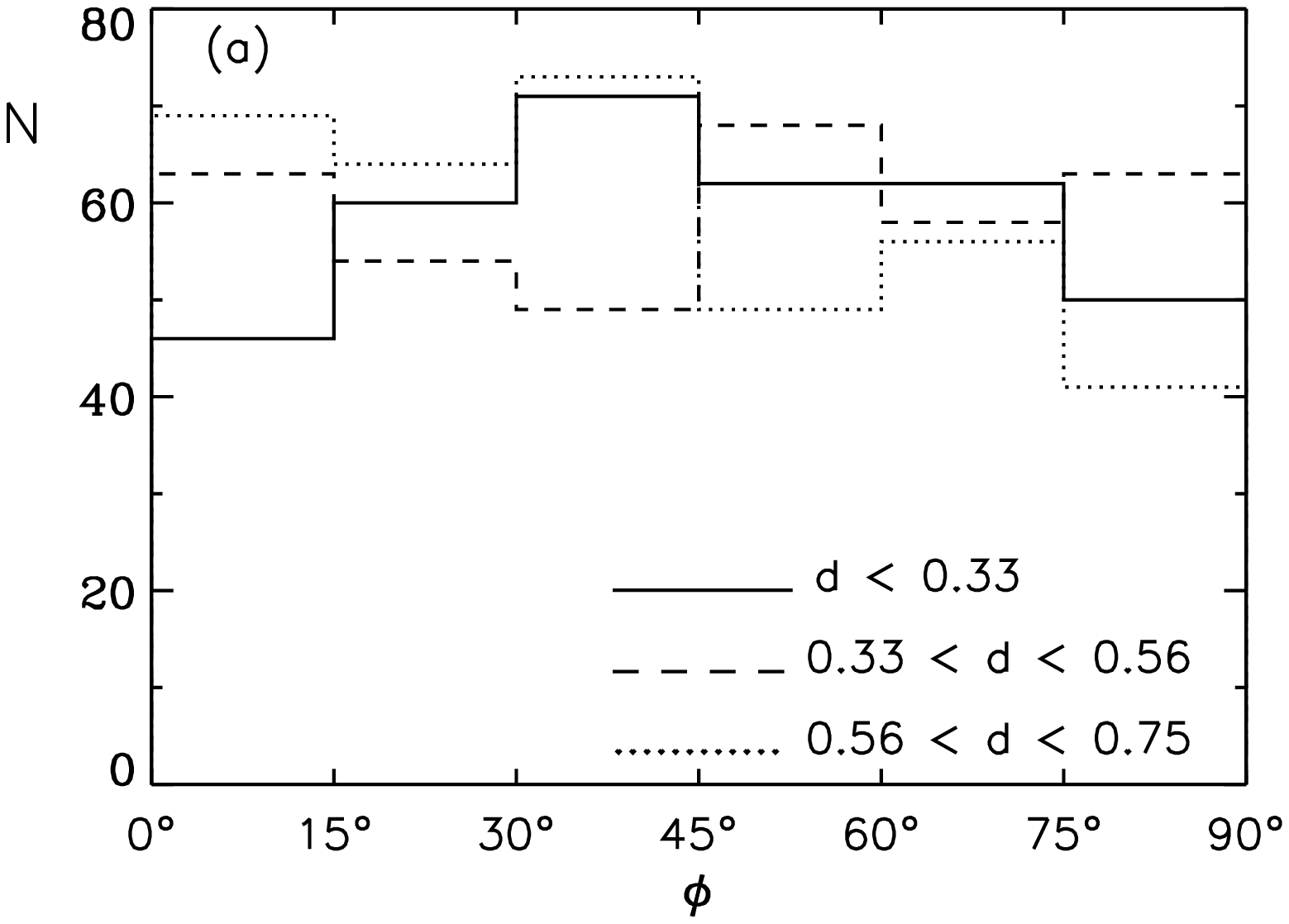,width=6.8cm}
\hspace*{-6mm}
\epsfig{file=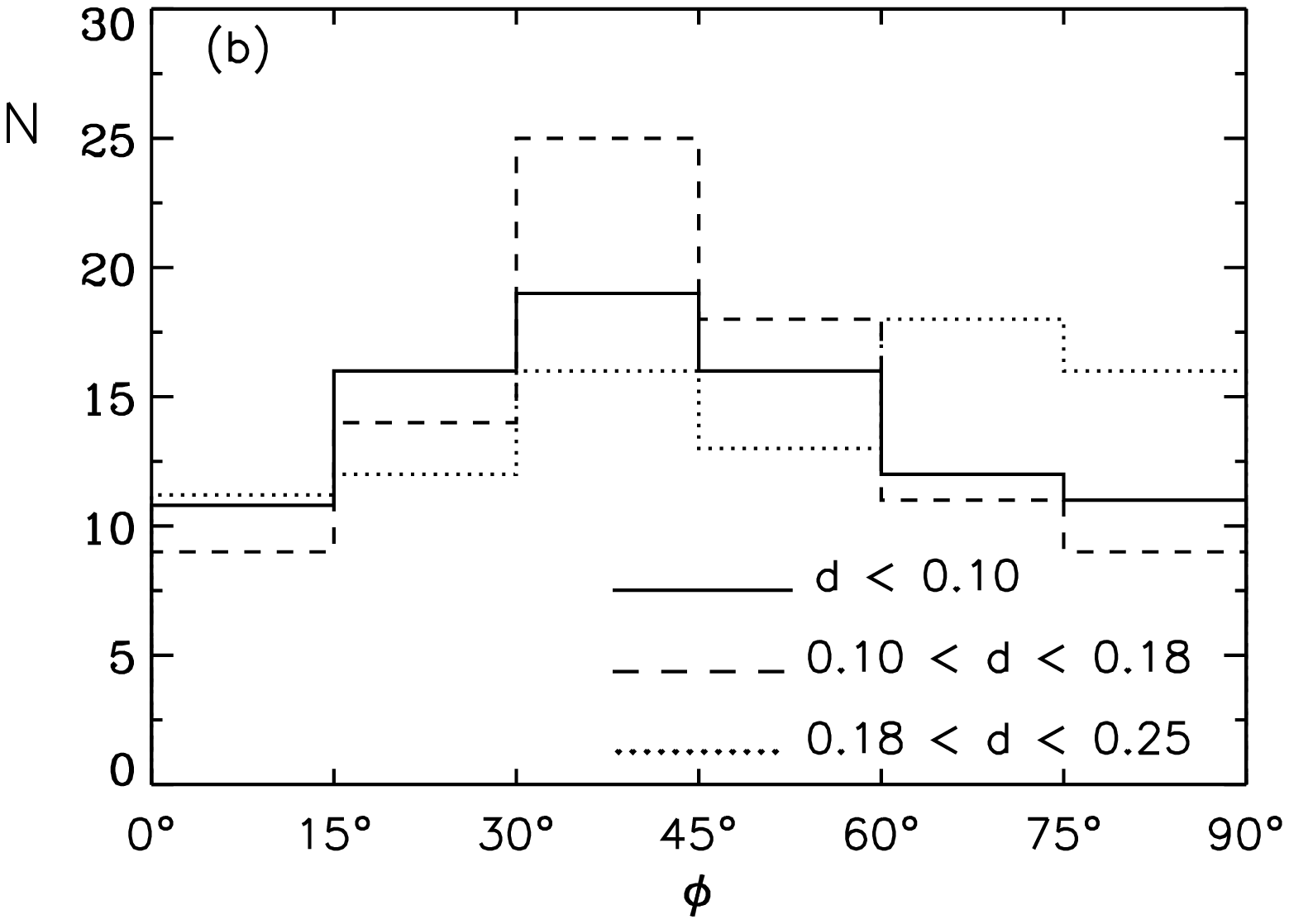,width=6.8cm}
}
\hspace*{-9mm}
\vspace*{-2mm}
\mbox{
\epsfig{file=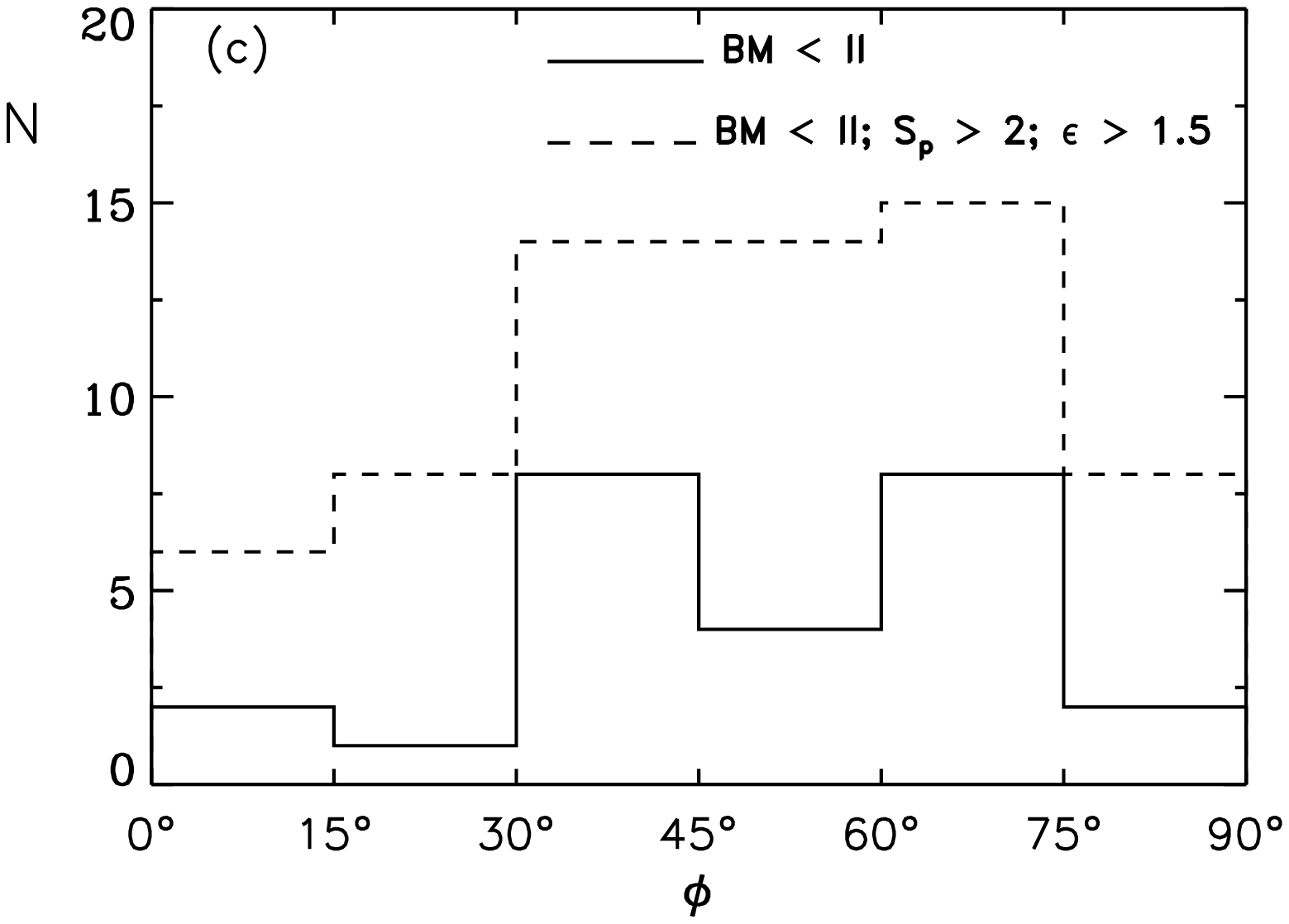,width=6.8cm}
\hspace*{-6mm}
\epsfig{file=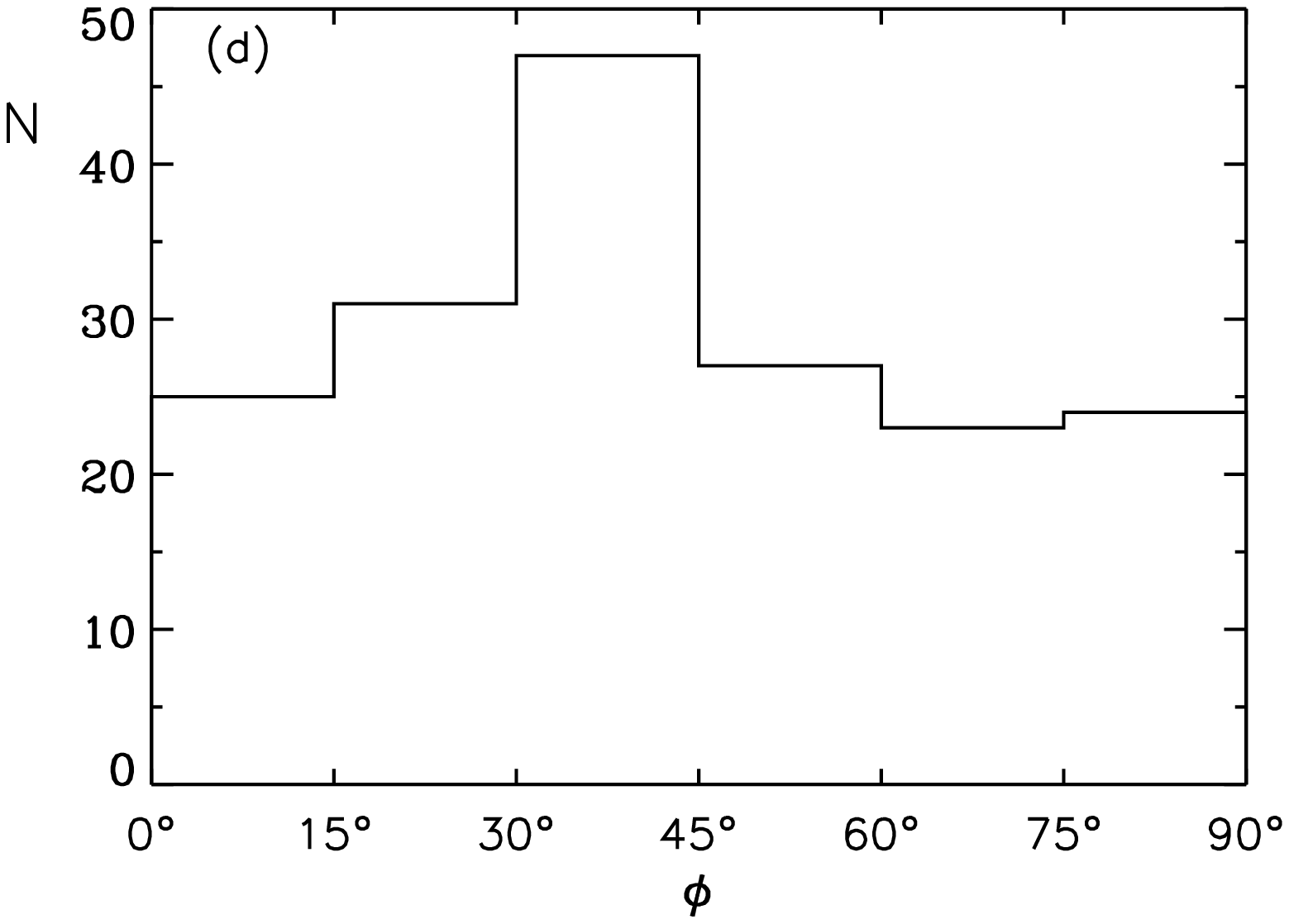,width=6.8cm}
}
\vspace{-6mm}  
\caption{Orientation angles of FIRST sources in ACO clusters:
{\bf a/b)} D$\ge$5, S$_{p}\ge$2\,mJy, $\epsilon\ge$1.5,
for d$_{\rm A}\le$0.75 (a) and d$_{\rm A}\le$0.25 (b), both split 
into 3 equal subsamples;
{\bf c)}  D$\ge$5, d$_{\rm A}\le$0.25, BM$<$II, with and without restricting 
S$_{p}$ and $\epsilon$; {\bf d)} D=[4,5], d$_{\rm A}\le$0.25, 
S$_{p}\ge$2\,mJy, $\epsilon\ge$1.5 (cf.\,text)}
\end{figure}

\section{Are there really no Radio Arcs in FIRST\,?}

Considering that the source sample will anyway be contaminated by many 
of the elongated, tailed, FR\,I type sources typically found in clusters,
we tried to find potential arc candidates by inspection of maps of
individual sources.
Using the cluster database maintained at Astronomical Institute of 
St.-Petersburg University \cite{GA97}
we examined the FIRST and Digitized Sky Survey (DSS) images
and APM charts    \cite{APM} for two subsamples.
We looked at all 42 FIRST sources
with d$_{\rm A}<$0.2, D$\ge$5, S$_{p}\ge$2\,mJy, $\epsilon\ge$1.5, 
and $\phi\ge$70$^{\circ}$. One of these sources (4C\,39.29) 
stands out due to its high flux (S$_{1.5}$=1.4\,Jy). It is located 2.7$'$
(d$_{\rm A}$=0.3) SE of the cD in A\,963, a rich X-ray cluster 
at z$_{\rm cl}$=0.21, which is known for its giant optical 
arcs within $\sim$20$''$ N and S of the cD \cite{LavHen88}. 
Unpublished high-resolution VLA maps \cite{LOH93} suggest
independently that the source may be lensed. 
The next strongest candidate (4\,$\times$ weaker)
is a wide-angle-tail (WAT) source close to the X-ray centre of A1190, 
previously mapped by \cite{OO85}. 
The FIRST and DSS images for this and several other 
sources revealed that either the radio PAs refer to subcomponents
of more complex sources, or that $\phi\ge$70$^{\circ}$ is due
only to poorly defined Abell centres, or both.

Two examples are given in the upper panels of Fig.\,2:
a 3C\,465-type WAT source in A1438 (z$_{\rm est}$=.16, R=1, BM=III),
and a complex 
WAT in A2110 (z$_{\rm cl}$=0.1, R=1, BM=I-II), 
probably seen nearly end-on. Both have prominent parent galaxies.

\begin{figure}
\hspace*{-24mm}
\vspace{-10mm}  
\mbox{
\epsfig{file=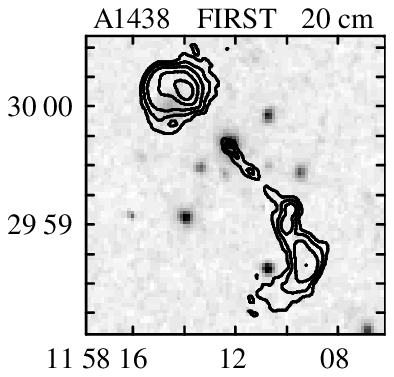,width=8.3cm,angle=270}
\hspace*{-19mm}
\epsfig{file=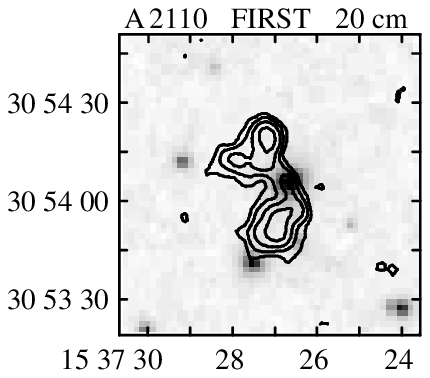,width=8.3cm,angle=270}
}
\hspace*{-24mm}
\vspace*{-14mm}
\mbox{
\epsfig{file=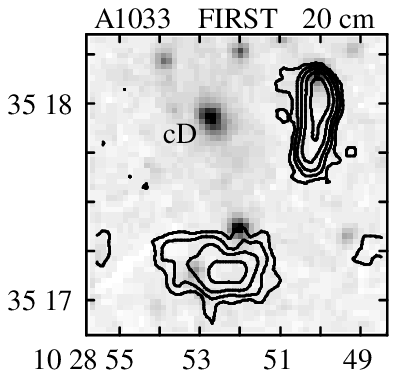,width=8.3cm,angle=270}
\hspace*{-19mm}
\epsfig{file=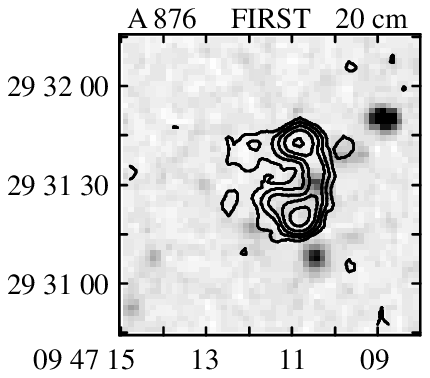,width=8.3cm,angle=270}
}
\caption{FIRST sources near ACO centres: radio contours and optical DSS 
in greyscale (B1950); except for A1033 we found no previously published 
radio maps (see text)}
\end{figure} 

We also inspected about 150 sources in 65 clusters with R$\ge$2 and 
z$>$0.1 which may be more likely to contain radio arcs.
We determined improved cluster centres from DSS images,
but found no preference for a ``tangential'' orientation 
($\phi>$45$^{\circ}$) of unidentified (presumably background) radio 
sources around these revised centres.
Moreover, we find that most of the sources seen towards the
cluster cores can be identified with likely cluster members or other 
galaxies, leaving a very poor statistics for the 
remaining unidentified radio sources. We note
that \cite{BK95} had only excluded the cD galaxies as cluster
members from their analysis.

The lower panels of Fig.\,2 show two sources with intriguing shapes.
The cD galaxy in the southern clump of the rich (R=2, BM=III) X-ray \cite{All92} 
cluster A1033 (z$_{\rm cl}$=0.1) is $\sim3'$ (d$_{\rm A}$=0.22)
SE from the ACO centre, 
much further than the characteristic Einstein radius ($\sim$45$''$). 
Close to the cD (which is also the X-ray centre)
there are two FIRST radio sources, both having their 
major axes perpendicular to the radius vector from the cD.
A much deeper 1.4\,GHz map \cite{LawGre95} shows
the arc-like source due W to be a head-tail source in the cluster.  
However, its high surface brightness and abrupt termination 
of the tail are unusual.
The southern source has a very steep spectrum and lacks an optical 
identification. It has been interpreted as a cluster radio halo  
\cite{LawGre95}, but with its 100$''\times$10$''$ arc-like shape 
\cite{LawGre95} it is equally unusual and merits further study.
The source in A\,876 (z$_{\rm est}$=.17, R=1, BM=II:) drew our attention due 
to a radio ``arc'' connecting two compact sources $\sim$25$''$ apart. 
However, the ``arc'' is curved away from the cluster centre and 
almost certainly due to
the inner, smoothly-curved jets of a WAT source coinciding 
with a likely cluster member (m$^{R}_{\rm APM}$=17.3).

\section{Discussion}

The typical offsets of Abell centres from the true centres of mass
are usually larger than typical Einstein radii. 
This is true at least for distant clusters and particularly
if we consider the subcluster centres of F-type and other clusters 
as distinct cluster cores. Thus, poor Abell centres alone prevent us
from finding statistical GL,
and extending the same statistical analysis to a larger number of clusters 
would not remove this obstacle. 
Our analysis of individual objects proves that a statistical source
selection by catalog parameters (like BM, RS, d$_{\rm A}$, $\phi$, 
$\epsilon$, etc.) is insufficient for finding good arc candidates. 

An inspection of FIRST and DSS images of large numbers of sources near
accurately known cluster centres may discover one or two radio arcs;
however, the theoretical predictions indicate that we need an order of
magnitude increase in brightness sensitivity over that of FIRST before the
number of radio arcs can approach those discovered in the optical domain.
This is best achieved by deep radio mapping of small samples of distant,
X-ray luminous clusters with accurate centres of mass determined from 
X-ray maps. \\[-1.5ex]


We thank R.L.\,White for the FIRST map server, A.\,Fletcher and J.\,Wambsganss
for useful comments, N.\,Loiseau for Fig.\,1, and E.\,Brinks
for help with Fig.\,2. {\tt SkyView}, APM, and NED did good jobs, too. 
H.A.~received a travel grant from the meeting organizers, 
and A.G. financial support from RFBR (grant N 97-02-18212). \\[-4.ex]

{}

\end{document}